# ArXiV 1606.01960/physics.ins-det

**Advanced broad-band solid-state supermirror polarizers for cold neutrons**

A.K. Petukhov, V.V. Nesvizhevsky*, T. Bigault, P. Courtois, D. Jullien, T. Soldner

*Institut Max von Laue – Paul Langevin, 71 avenue des Martyrs, F-38042, Grenoble, France ;*
petukhov@ill.fr, nesvizhevsky@ill.eu, bigault@ill.fr, courtois@ill.fr, djullien@ill.fr, soldner@ill.fr

**Abstract**

An ideal solid-state supermirror (SM) neutron polarizer assumes total reflection of neutrons from the SM coating for one spin-component and total absorption for the other, thus providing a perfectly polarized neutron beam at the exit. However, in practice, the substrate's neutron-nuclei optical potential does not match perfectly that for spin-down neutrons in the SM. For a positive step in the optical potential (as in a $Fe/SiN_x$ SM on $Si$ substrate), this mismatch results in spin-independent total reflection for neutrons with small momentum transfer $Q$, limiting the useful neutron bandwidth in the low-$Q$ region. To overcome this limitation, we propose to replace $Si$ single-crystal substrates by media with higher optical potential than that for spin-down neutrons in the SM ferromagnetic layers. We found single-crystal sapphire and single-crystal quartz as good candidates for solid-state $Fe/SiN_x$ SM polarizers. To verify this idea, we coated a thick plate of single-crystal sapphire with a $m = 2.4$ $Fe/SiN_x$ SM. At the T3 instrument at the ILL, we measured the spin-up and spin-down reflectivity curves with $\lambda = 7.5$ Å neutrons incident from the substrate to the interface between the substrate and the SM coating. The results of this experimental test were in excellent agreement with our expectations: the bandwidth of high polarizing power extended significantly into the low-$Q$ region. This finding, together with the possibility to apply a strong magnetizing field, opens a new road to produce high-efficient solid-state SM polarizers with an extended neutron wavelength bandwidth and near-to-perfect polarizing power.

*Keywords: Neutron, super-mirror, polarizing bender, solid-state polarizer*

## 1. Introduction

The only known technique to polarize cold neutrons of a broad wavelength band ($\lambda \in [2 - 20 \text{ Å}]$), at acceptable loss of intensity, is their reflection/transmission from/through multi-layer structures called "Super-Mirrors" (SMs) [1-5]. Multilayer structures have found broad application in neutron instrumentation, for instance as monochromators, polarizers, spectrum shaping and focusing devices.

The interaction of cold neutrons with matter in the s-wave approximation can be described in terms of the effective Fermi pseudo-potential, also known as optical potential:

$$U_n = \frac{2\pi\hbar^2}{m_n}\sum_j N_j b_j = \frac{2\pi\hbar^2}{m_n}\rho, \qquad (1)$$

where $\hbar$ is the reduced Planck constant, $m_n$ the neutron mass, $b_j$ the neutron scattering length, $N_j$ the number density of nuclei, and $\rho = \sum_j N_j b_j$ the scattering length density, SLD; the summation runs over all elements and isotopes that constitute the layer. Neutron waves propagating through a multilayer undergo multiple reflections at interfaces and the resulting reflectivity and transmittance of the structure are determined by the interference of all reflected and transmitted waves. The interference pattern depends on the phases of the summed waves, and thus on the layer thicknesses and the amplitudes of the neutron wave vectors in the layers. The latter ones are defined by the following expression:

$$k = \sqrt{\frac{2m_n}{\hbar^2}(E_0 - U_n)}, \qquad (2)$$

where $E_0$ is the energy of an incident neutron in vacuum. For a one-dimensional potential structure in the direction normal to the interface, the $k_\parallel$ component is constant and only the $k_\perp$ component varies:

$$k_\perp = \sqrt{\frac{2m_n}{\hbar^2}(E_\perp - U_n)}. \qquad (3)$$

Here $E_\perp = \frac{k_{0\perp}^2}{2m_n}$ is the part of energy in vacuum associated with the normal momentum component $k_{0\perp}$. The reflection coefficient $R$ from the interface between vacuum and semi-infinite matter reads [6]:

$$R = \left|\frac{k_{0\perp} - k_\perp}{k_{0\perp} + k_\perp}\right|^2 = \left|\frac{1 - \sqrt{1 - \frac{\lambda^2}{\pi \sin^2(\Theta)}\rho}}{1 + \sqrt{1 + \frac{\lambda^2}{\pi \sin^2(\Theta)}\rho}}\right|^2. \qquad (4)$$

In general, a SM is designed as an aperiodic multilayer sequence, which provides neutron reflection at small grazing angles. An important property of a SM is its critical momentum transfer, $Q_c^{SM}$; up to this value of momentum transfer the SM reflects neutrons. By convention, it is measured in multiples, $m$, of the natural $Ni$ critical momentum transfer $Q_c^{Ni}$, the largest critical momentum transfer among naturally occurring elements:

$$Q_c^{SM} = mQ_c^{Ni}, \qquad (5)$$

$$Q_c^{Ni} = 4\sqrt{\pi\rho_{Ni}}. \qquad (6)$$

If a reflecting material is magnetic, the effective optical potential $U$ also includes the interaction of the neutron magnetic moment with the magnetic field $B$:

$$U = U_n \mp \mu_n B, \tag{7}$$

where $\mu_n \approx -60\ neV/T$ is the neutron magnetic moment. The interaction of the neutron magnetic moment with the material magnetic field can be accounted for by introducing a magnetic scattering length:

$$\rho_m = \frac{m_n}{2\pi\hbar^2} |\mu_n B|, \tag{8}$$

$$\rho^\pm = \rho_n \pm \rho_m, \tag{9}$$

where the upper/lower sign corresponds to the neutron spin parallel/antiparallel to the magnetic field direction. The difference (9) in SLD results in different critical angles for the two spin components:

$$Sin(\Theta_c^\pm) = \lambda\sqrt{\rho^\pm/\pi}. \tag{10}$$

This effect opens naturally a simple way to spin-polarize neutrons by reflecting them from magnetically saturated materials (the saturation field in $Fe$ is $\approx 2.2\ T$).

2. **Design of polarizing bender**

Since the first invention [1], a variety of mirror and SM benders and devices have been proposed and actually constructed to provide polarized neutron beams [7-14]. A comparative description of several configurations of polarizing SM elements is given in ref. [13]. To polarize cold neutron beams with large cross-section and large angular divergency, the most powerful tool is the polarizing bender. A typical design of a polarizing bender is shown in Fig. 1; an example [15] is the $Co/Ti$ polarizer successfully operating already for more than 10 years at the PF1B instrument at the ILL [16].

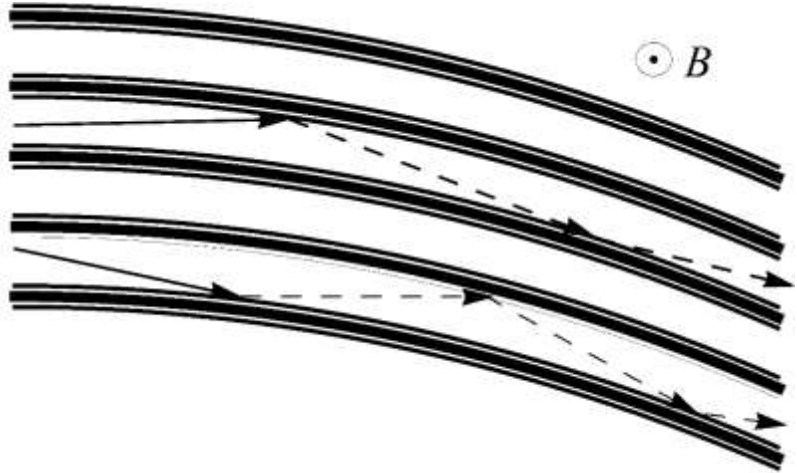

Fig. 1. The cold neutron polarizer at PF1B: 30 channels of 80 cm length, air gaps of $2\ mm$, borated float glass substrates $0.74\ mm$ thick, $m = 2.8$ CoTi SM coating, total cross-section $80 \times 80\ mm^2$, curvature radius $R = 30\ m$, applied magnetic field $120\ mT$, polarization averaged over the transmitted neutron capture flux at PF1B $98.5\ \%$, transmission for unpolarized neutrons $24\ \%$.

A classical polarizing bender is assembled in a stack of well-polished (typical rms roughness $R_a <$ 0.5 $nm$) substrates coated on both sides with spin-polarizing SMs and separated with air gaps (often with additional anti-reflecting and absorbing layers), see Fig. 1. The whole assembly is curved to provide at least one neutron collision with reflecting plates. In devices of this type, neutrons travel through channels formed by neighbouring reflecting plates. SM materials are chosen to provide high reflectivity for one spin-component and low reflectivity for the other spin-component. Combinations $Co/Ti$, $FeCo - TiZr$ [17] and $Fe_{50}Co_{48}V_2 - TiN_x$ [18] are commonly used in air-gapped SM polarizers.

With the availability of commercial single-crystal $Si$ wafers, and the progress in SM technology, the past decade has seen the development of a new family of neutron optical elements, where the transmitting medium is $Si$ instead of air: so-called "solid state" neutron optical elements [19-20]. In particular, polarizing benders have largely benefited from this progress, from the first device [3] to more common ones [21-22]. SM materials for this type of polarizers are chosen to provide SLDs for spin-down neutrons close to the single-crystal $Si$ SLD. Combinations $Fe/SiN_x$ [23] and $Fe_{89}Co_{11}/Si$ are commonly used in solid-state polarizers. For solid-state polarizer substrates, it is crucial to select a single-crystal material (to avoid Bragg reflection from small crystalline domains) with very low neutron absorption. Single crystal $Si$ wafers available on the market meet these requirements.

Presently, solid $Fe/SiN_x$ SM polarizing benders appear to be the most attractive solution since besides their compactness (the length of solid SM polarizers is typically ≈10 times smaller than the length of air gap polarizing benders), they also allow avoiding problems associated with the long-life ($T_{1/2} \approx$ 5.27 $years$) isotope $^{60}Co$ produced via neutron capture.

### 3. Polarizing mirror design

#### 3.1 From classical to advanced design

In an ideal polarizer, one spin component propagates through the SM without reflection and then is absorbed in a capping $Gd$ layer. The other component is fully reflected from the SM, see Fig. 2. This behavior requires perfect matching of the SLDs of the SM materials for one spin component and high contrast for the other one. In practice, materials used in SMs never match perfectly. Materials of interest for neutron polarizing applications are listed in Table 1.

|  | $Fe$ | $Co$ | $Ni$ | $Ti$ | $Si$ | $SiO_2$ quartz | $Al_2O_3$ sapphire | $Gd$ |
|---|---|---|---|---|---|---|---|---|
| $\rho_n$ | 8.02 | 2.27 | 9.4 | -1.93 | 2.08 | 4.19 | 5.72 | 2.24-0.325i |
| $\rho^+$ | 12.97 | 6.38 | 10.86 |  |  |  |  |  |
| $\rho^-$ | 3.08 | -2.14 | 7.94 |  |  |  |  |  |

*Table 1. Neutron scattering length density, in $10^{-6}$ Å$^{-2}$, of various bender materials* [24].

The reflection probabilities at each interface are governed by the difference between the SLD $\rho$ of the nonmagnetic layer ($SiN_x$) and either $\rho^+$ or $\rho^-$ of the ferromagnetic layer ($Fe$), depending on the neutron's spin.

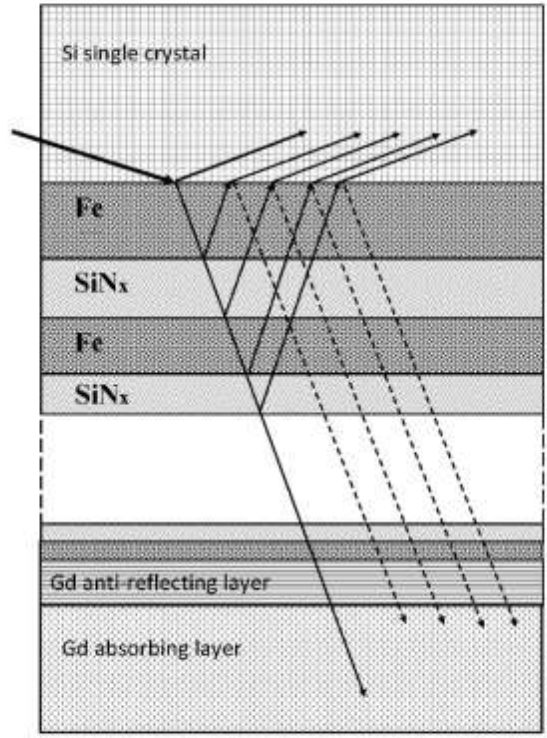

*Fig. 2. A design of $Fe/SiN_x$ SM coating for solid state polarizers.*

As follows from Table 1, the SLD of $Si$ ($\rho_{Si} = 2.08 \cdot 10^{-6}$ Å$^{-2}$) is lower than the SLD of $Fe$ for spin-down neutrons ($\rho_{Fe}^- = 3.08 \cdot 10^{-6}$ Å$^{-2}$) resulting in unfavorable total reflection for spin-down neutrons:

$$Sin(\Theta_c^-) = \lambda \sqrt{\frac{\rho_{Fe}^- - \rho_{Si}}{\pi}}. \quad (11)$$

Another effect of this mismatch is a low-contrast modulation of the SLD profile for spin-down neutrons with the SM sequence. Fortunately, the extra reflectivity caused by this modulation can be significantly diminished using so-called reactive magnetron deposition of $Si$ in the presence of $N_2$ gas flow [23]. This technique provides $SiN_x$ compound layers instead of pure $Si$ layers; the SLD of the $SiN_x$ layers depends on $N_2$ gas flow. At an optimum flow of $N_2$, the SLD of $SiN_x$ can be almost perfectly matched with $\rho_{Fe}^-$.

The position of the high-$Q$ edge of the spin-up neutron reflectivity is controlled by the layer of the smallest thickness while the low-$Q$ behavior of the spin-down reflectivity depends mainly on the properties of the thickest layer in the SM sequence. With a good accuracy, the reflection in the low-$Q$ region for a $FeSiN_x$ SM reads:

$$R^{\pm} \approx \left| \frac{\left(1-\sqrt{1+\frac{\lambda^2}{\pi Sin^2\Theta}(\rho_{Si}-\rho_{Fe}^{\pm})}\right)}{\left(1+\sqrt{1+\frac{\lambda^2}{\pi Sin^2\Theta}(\rho_{Si}-\rho_{Fe}^{\pm})}\right)} \right|^2, \qquad (12)$$

and the neutron polarization after a single reflection is given by

$$P = \frac{R^+ - R^-}{R^+ + R^-}. \qquad (13)$$

Usually, in good polarizers $R^+$ is close to unity and $R^- \ll R^+$. The reflectivity $R^+$ controls mainly the transmission efficiency while $R^-$ is responsible for the polarizing power. From (12) we notice that the reflectivity at the interface between the non-magnetic substrate and the thickest ferro-magnetic layer ($Fe$) is very different for positive and negative steps in SLD, see Fig. 3.

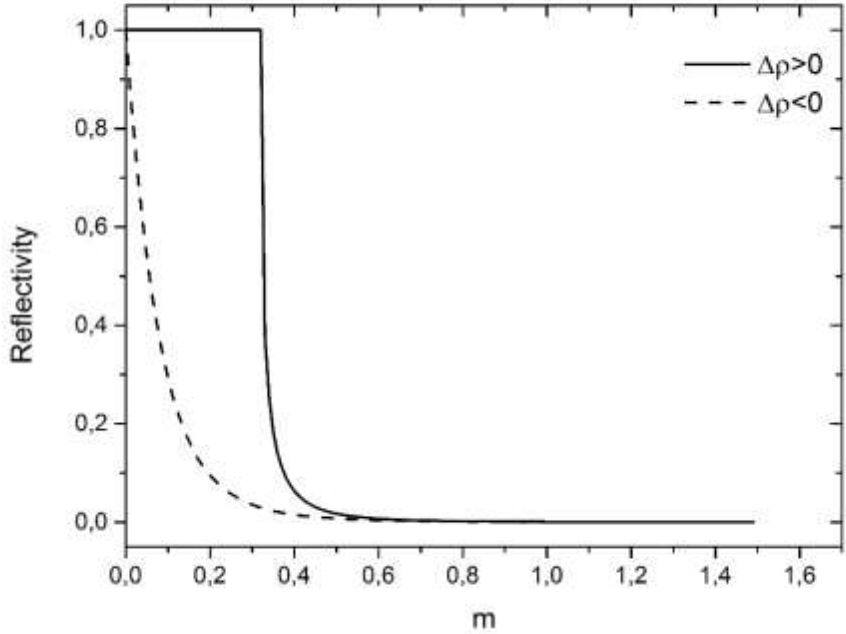

*Fig. 3. Calculated reflectivity as a function of momentum transfer (in units of $Q_c^{Ni}$) for spin-down neutrons reflected at the interface between substrate and a thick layer of $Fe$ for positive and negative steps in SLD. The amplitude of the step is the same: $|\Delta\rho| = 10^{-6} Å^{-2}$.*

Reflection at a positive step in SLD (solid line) results in the regime of total reflection in the low-$Q$ region, see Eq. (11), while reflection at a negative step in SLD (dashed line) is much weaker and does not show a total reflection regime. The smaller the step magnitude, the sharper the decrease of reflectivity with increasing grazing angle. We conclude that a negative SLD step seen by spin-down neutrons is better than a positive one, for good polarizing mirrors. In case of solid-state SM polarizers, this condition favors substrates with a SLD higher than the SLD for spin-down neutrons in the ferro-magnetic layer of the SM coating. Another requirement is very high transparency for neutrons propagating through the substrate material. Single crystal $Si$ wafers commonly used as substrates for solid-state $Fe/SiN_x$ polarizers meet this requirement but not that of a negative step in SLD. Therefore, they lead to a mediocre neutron wavelength bandwidth of polarizing devices. Better SLD matching for spin-down neutrons can be achieved by replacing pure $Fe$ by the $Fe_{89}Co_{11}$ alloy. However, such a solution is acceptable only if the device is going to be used in low incident neutron fluxes, otherwise $Co$ will be highly activated. As follows from Table 1, good candidates for the substrate material, which satisfy both requirements, are single-crystal quartz and single-crystal sapphire wafers. Both materials are very transparent for cold neutrons and have SLDs higher than $\rho^-$ for $Fe$.

### 3.2 Reflectivity calculations

The main conclusions of the previous section are based on a simplified model of reflection at the interface between two semi-infinite materials; this model works well in the low-$Q$ region. In this section, we compare predictions within a more realistic model. Fig. 4 presents a deposition scheme for a solid $m = 3.2$ supermirror for neutrons incident from inside the substrate using the "inverted" $Fe/SiN_x$ multilayer sequence. The term "inverted" multilayer sequence refers to the sequence of sputtering: The sputtering starts with the thickest layer which is inverted compared to SMs for neutrons incident from air. This means that neutrons incident on the SM interface from inside the substrate see first the thickest layer in the multilayer sequence.

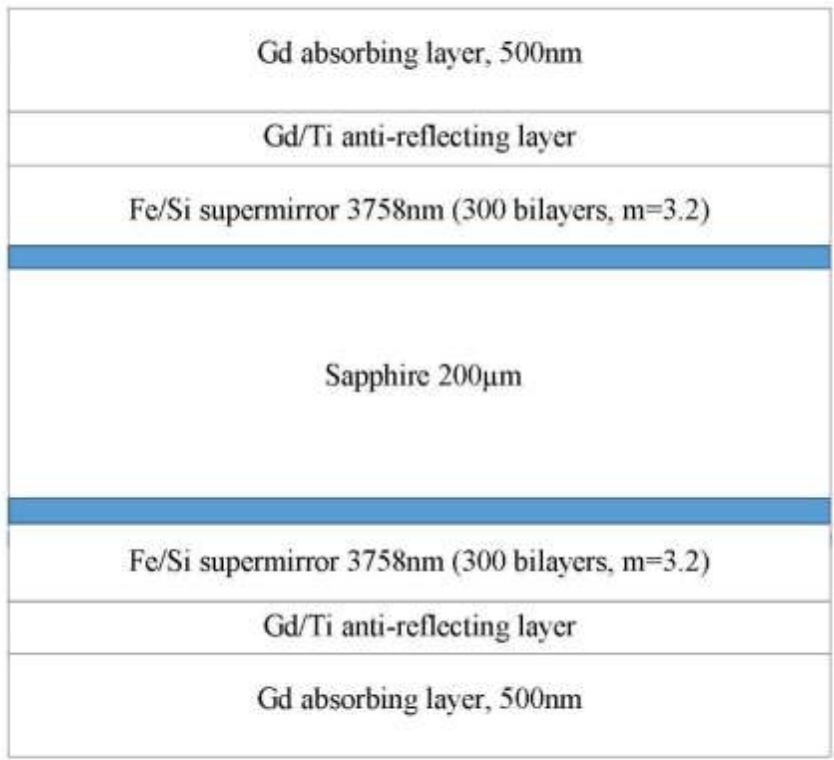

*Fig. 4. A deposition scheme.*

Fig. 5 depicts calculated spin-dependent reflectivity curves for neutrons incident on a SM interface from inside the substrate for different materials of the substrate ($Si$, quartz, sapphire). The reflectivity curves are calculated using the IMD package [25-26]. As expected, the spin-up reflectivity is nearly identical (in practice, high-frequency oscillations are fully washed out by the angular divergence of incident neutrons). A difference can be clearly observed for the spin-down reflectivity curves (gray curves). No total reflection for spin-down neutrons is observed for single-crystal sapphire and quartz substrates. For single crystal $Si$ substrates, total reflection is present (at $m < 0.4$) even for spin-down neutrons, due to the positive step in SLD, see Table 1. All spin-down reflectivity curves show a broad minimum at $0.75 < m < 1.5$. For higher $m$, the spin-down reflectivity increases gradually due to the high-frequency modulation of the $\rho^-$ profile by a "non-magnetic dead-layers" formation in the interface between $Si$ and $Fe$ [27]. It is the "dead-layer" effect that limits polarization in a single reflection. We found that for all three substrates the polarization expected after a single reflection is nearly the same for high-$Q$ reflections ($m > 0.75$). For the low-$Q$ region ($m < 0.75$), polarization depends strongly on the substrate material. It is the worst for the $Si$ substrate (black points) and much better for sapphire and quartz (open triangles and squares, respectively).

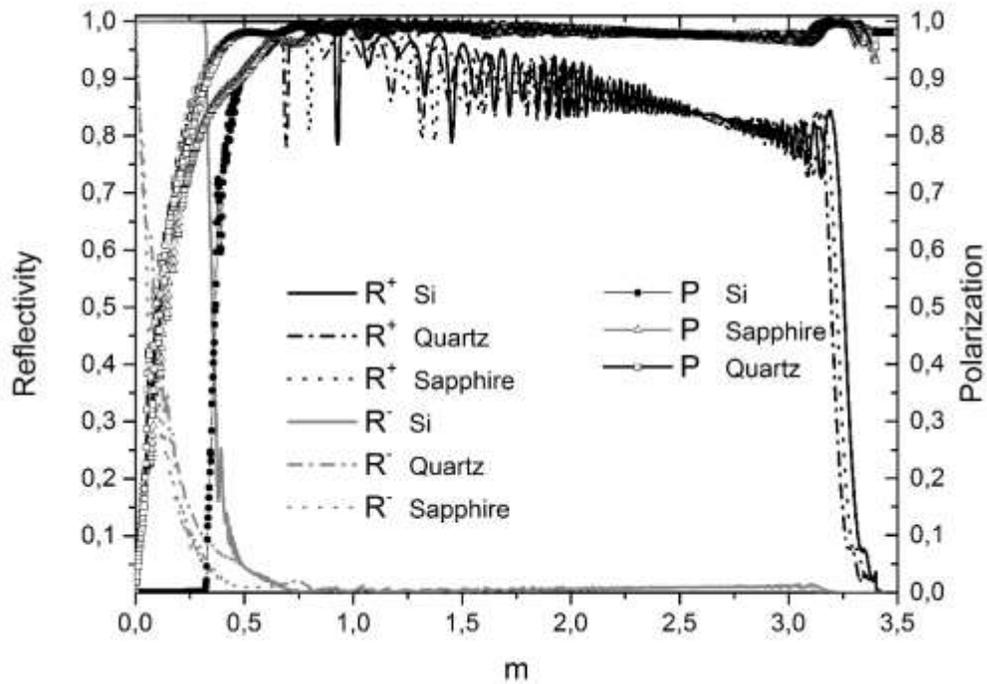

Fig. 5. Simulated spin-dependent ($R^+$ - black lines, $R^-$ - gray lines) reflectivity for $Fe/SiN_x$ SM deposited on different substrate materials. The symbols (black dots for $Si$ substrate, open triangles for sapphire substrate, open squares for quartz) represent the estimated polarization after a single reflection. Neutrons are incident on the interface from the substrate.

4. **Measurements**

Using our "in house" magnetron sputtering facilities, we produced two samples: a thick (12.5 $mm$) optically polished single-crystal sapphire plate (diameter 65 $mm$) coated with a $m = 2.5$ $Fe/SiN_x$ SM and a thick (10 $mm$) polished (rms roughness $R_a < 0.5\ nm$) single-crystal $Si$ plate (80 $mm$ along the beam) with the same coating. Spin-dependent reflectivity curves for both samples were measured using the ILL test instrument T3 (wavelength of incident neutrons $\lambda = 7.5$ Å, beam width at the sample position ≈ 0.08 $mm$, beam angular divergence ±0.15 $mrad$, flipper efficiency ≈ 99 %, incident neutron beam polarization ≥ 99 %, magnetic field strength at the sample position 60 $mT$).

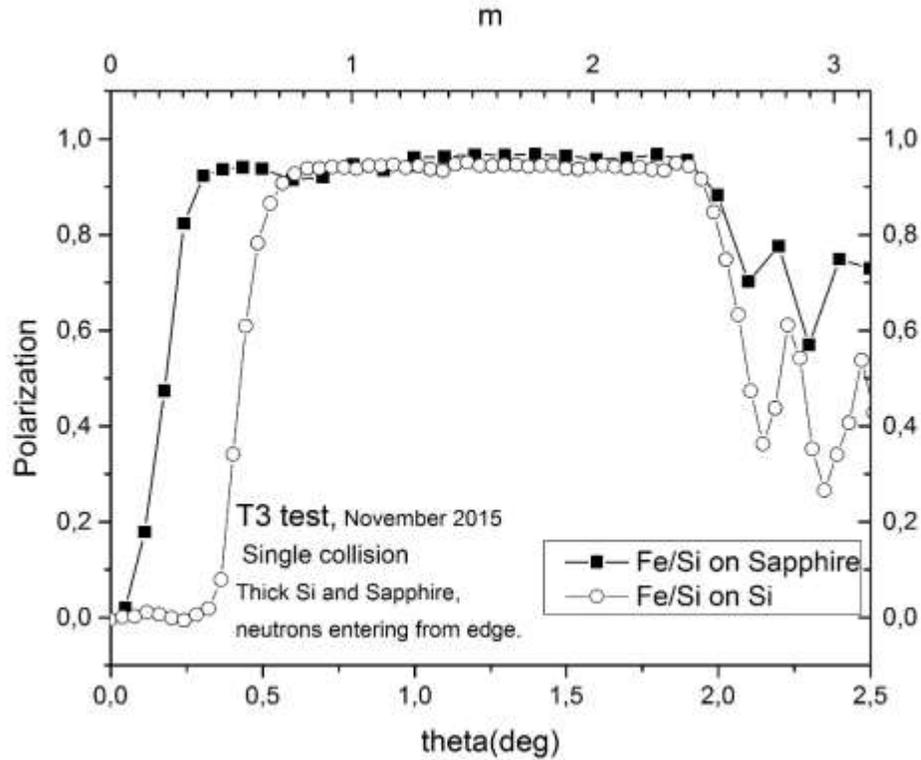

*Fig. 6. Measured neutron polarization after a single reflection at a $Fe/SiN_x$ SM coating deposited on single crystal sapphire (black points) and on single crystal Si (open points). No corrections for the instrumental polarizing and flipping imperfections, amounting to slightly above one percent, are done. Neutrons were incident to the SM interface from inside the substrate.*

For sample reflectivity measurements in the region of very small angles (well below one degree), geometrical correction for the sample finite size is very important. In Fig. 6, we depict the SM polarizing power defined according to Eq. (13) (contrary to the reflectivity, the polarizing power does not need a geometrical correction). As expected, the positions of the high-$Q$ cut-off in the high polarizing power window are practically independent on the substrate material whereas the position of the low-$Q$ cut-off in polarization is significantly lower for the sapphire substrate indicating the absence of total reflection for spin-down neutrons incident from inside the substrate on the $Fe/SiN_x$ interface.

5. **C-bender simulations**

Using the calculated reflectivity curves described in section 3.2, we simulate the performance of a solid-state bender with $m = 3.2$ $Fe/SiN_x$ SM built using substrates of different materials (quartz, sapphire, $Si$). For the simulation, we use an "in-house" ray-tracing package written in "Mathematica". We assume a perfect geometry of rectangular plates (50 $mm$ length, 0.2 $mm$ thick) curved with the radius of $R = 1.25\ m$ corresponding to the cut-off wavelength $\lambda_{cut-off} = 0.37\ nm$, defined as

$$\lambda_{cut-off} = \frac{\gamma}{2}/(m\Theta_c^{Ni}), \tag{14}$$

with the bending angle $\gamma$ (21). We also assume that the bender is installed behind a main neutron guide with $m = 1$ angular divergence at the exit. Another important assumption is complete magnetic saturation of the ferro-magnetic layer meaning that the spin-flip reflectivity is set to be zero: $R_{+-} = R_{-+} = 0$. As shown in [14, 28], this assumption is valid only for a very high magnetizing field $B \geq 0.3\ T$ (approximately 10 times higher than the typical field $B = 30 - 50\ mT$ used to magnetize SMs). For a magnetic field of low strength, $B \leq 0.1\ T$, a significant depolarization effect $\Delta P \geq 1\ \%$ can be observed. The magnitude of depolarization depends on the momentum transfer $Q$, on the $m$-value of the SM, on the SM composition, and on the strength of the applied field [28]. Applying a very high magnetic field of $B \geq 0.3\ T$ to an air-gap SM bender is challenging, if possible. On the other hand, this is definitely possible for a very compact solid-state SM polarizing bender.

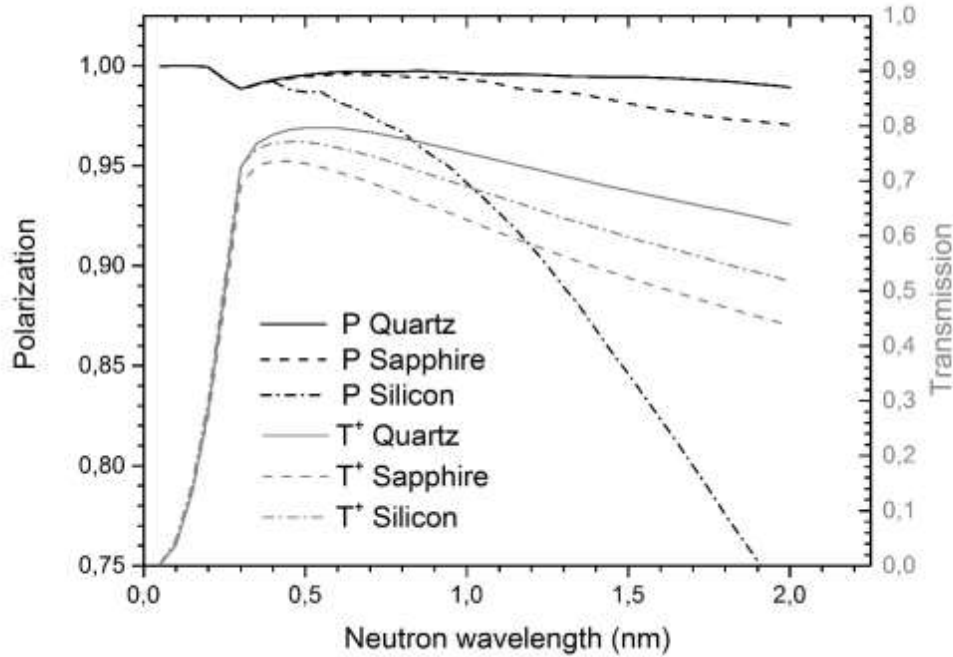

*Fig. 7. Simulated polarization (black lines) and transmission of spin-up neutrons (grey lines) after a solid-state C-bender for different substrate materials coated with $Fe/SiN_x$ SM.*

Fig. 7 shows results of our simulations for the solid C-bender described above for different substrate materials (quartz, sapphire, $Si$). The black lines represent the polarizing power of the polarizer and the grey lines the transmission for spin-up neutrons. As expected, the absence of total reflection for sapphire and quartz improves dramatically the polarizing power for long-wavelength neutrons. For the quartz substrate with a smaller negative step in SLD at the substrate-SM interface, the polarizing power curve (solid) is practically flat and stays very close to 100 %. The next in quality of polarization is achieved with the sapphire substrate (dashed curve). The drop of polarization at the wavelength $\lambda = 2\ nm$ is about 3 % only. The worst result is obtained for the commonly used $Si$ substrate; the drop in polarization is larger than 25 %. In terms of transmission, all materials show similar performance in the region $\lambda \in [0.4 - 0.8\ nm]$. For longer wavelengths, the quartz substrate shows higher transmission due to its lower absorption cross-section. The difference in diffuse scattering cross-section (the dominant effect for short-wavelength neutrons [29]) for these materials plays a minor role since the transmission for short-wavelength neutrons is mainly governed by the short-wavelength cut-off Eq. (14) caused by the bender curvature. The small dip in polarization near the wavelength $\lambda \approx 0.3\ nm$ is caused by the "dead-layers" effect.

## 6. Conclusion

Basing on our analysis of limiting factors for the polarization in solid-state SM polarizers, and also on results of our very first experimental test, we formulate a new method to suppress unfavorable reflectivity for spin-down neutrons in the low-$Q$ range: Replacement of single-crystal $Si$ substrates by other materials with low absorption and a SLD higher than that for spin-down neutrons in the SM ferro-magnetic material. The negative step in SLD eliminates total reflection for spin-down neutrons in the low-$Q$ region. This low-$Q$ expansion of the window of high polarizing power in the reflectivity curves results in a dramatic improvement of the polarizing performance for long-wavelength neutrons. For $Fe/SiN_x$ solid polarizers, good substrate candidates are single-crystal quartz ($\rho = 4.19 \cdot 10^{-6}\ Å^{-2}$) and single-crystal sapphire ($\rho = 5.72 \cdot 10^{-6}\ Å^{-2}$). Nowadays, due to the revolution in Light-Emitting-Diodes technology, thin sapphire wafers (thickness of $\sim 0.15\ mm$) of sufficiently large diameter and polishing quality are easily available at a price comparable to $Si$ wafers. Applying strong magnetizing fields, $B \geq 0.3\ T$, can push the polarizing power of solid-state benders from the presently common value of 98 % to the level of 99.9 % and above [14, 28]. This opens a new road in the production of high-efficient solid-state SM polarizers with extended wavelength bandwidth and near-to-perfect polarizing power. Quartz single-crystal wafers of necessary size are also available and promise even better performance, however, they are less common and more expensive.